\documentclass[apjl]{emulateapj}

\usepackage{amsxtra}
\usepackage{amsmath}
\usepackage{amssymb}
\usepackage{graphicx}
\usepackage{txfonts}
\usepackage{natbib}
\usepackage{color}

\newcommand{\T}{\ensuremath{\Delta t}}

\newcommand{\p}[2]{\ensuremath{p_{#1}( {#2} )}}
\newcommand{\w}[2]{\ensuremath{w_{#1}( {#2} )}}

\newcommand{\ov}{\ensuremath{b^{xy}}}

\newcommand{\sref}[1]{Section~\ref{#1}}
\newcommand{\eref}[1]{Eq.(\ref{#1})}
\newcommand{\fref}[1]{Fig.\ref{#1}}

\begin{document}
\title{Markov properties of the magnetic field in the quiet solar photosphere.}
\author{A.Y. Gorobets}
\author{J.M. Borrero}
\author{S. Berdyugina}
\affil{Kiepenheuer-Institut f\"{u}r Sonnenphysik Sch\"{o}neckstr. 6, D-79104 Freiburg, Germany}
\email{a.y.gorobets@kis.uni-freiburg.de}

\begin{abstract}
The observed magnetic field on the solar surface is characterized by a very complex spatial
and temporal behaviour. Although feature-tracking algorithms have allowed us to deepen our
understanding of this behaviour, subjectivity plays an important role in the identification,
tracking of such features. In this paper we study the temporal stochasticity of the magnetic field on the solar
surface \textit{without} relying neither on the concept of magnetic feature nor on subjective
assumptions about their identification and interaction. The analysis is applied to observations
of the magnetic field of the quiet solar photosphere carried out with the IMaX instrument
on-board the stratospheric balloon {\sc Sunrise}. We show that the joint probability distribution functions of
the longitudinal ($B_\parallel$) and transverse ($B_\perp$)
components of the magnetic field, as well as of the magnetic pressure ($B^2=B^2_\perp+B^2_\parallel$), verify the
necessary and sufficient condition for the Markov chains. Therefore we establish that the magnetic
field, as seen by IMaX with a resolution of 0.15\arcsec-0.18\arcsec and $33$~sec cadence,
can be considered as a memoryless temporal fluctuating quantity.
\end{abstract}

\keywords{convection -- Sun: granulation -- Sun: photosphere -- Sun: surface magnetism }

\maketitle

\section{Introduction}
\label{section:intro}

The observed photospheric magnetic field appears as distributed concentrations over the entire solar surface.
These concentrations are charactirized by a variety of magnetic features (i.e. elements) that span over a
huge range of spatial scales, from active regions down to small-scale mixed-polarity features of the quiet
Sun network and internetwork \citep{Stenflo_rev}. In the quiet Sun (hereafter referred to as QS),
the aforementioned elements possess magnetic fluxes of the order of $10^{18}-10^{19}$ Mx
\citep{MagnetoChemistry,Parnell-model,Solanki-rev}. These elements also show rich and complex
dynamics, both in time and space, and interact with each other in a variety of ways as a consequence
of the  constant motions of the underlying flow patterns (i.e. convective motions). The characterization
of the elements is of crucial importance for many research topic within solar physics, such as:
understanding the coupling between the different solar atmospheric layers \citep{LC1,LC2}, relation
between the magnetic flux budget and coronal heating \citep{CH1}, extrapolations towards the solar
corona \citep{wiegelmann2013}, inferring semi-empirical magneto-hydrostatic models of the corona
\citep{wiegelmann2015} and solar wind \citep{solarwind,semiempirical}, etcetera.\\

The evolution of the QS magnetic features is studied in terms of flux emergence, cancellation,
coalescence and fragmentation that give a certain intermittent distribution of fluxes
over the solar surface. The statistics of the flux distribution is described by the so-called
magnetochemistry \citep{MagnetoChemistry}. Methodologically, the magnetochemistry is based on
the identification and tracking of particular features \citep{DeForestI,DeForestII,DeForestIII,DeForestIV,Iida}.
A prominent progress in our understanding of the solar surface magnetism has been achieved by methods based
on feature tracking \citep[e.g. ][; and references therein]{Thorton}. However, a comparison
of the different feature-tracking algorithms \citep{DeForestI}, has showed that the characterization
of the features is strongly affected by the choice of the algorithm and the assumptions they make
\citep[see also][]{Parnell-PowerLaw}.\\

A key concern was voiced by \cite{DeForestIV}: "measurement of the behavior
of small magnetic features on the photosphere is limited, partly by the spatial and temporal
resolution of the observing instruments, and partly by the difficulty of following visual features
that do not behave exactly like discrete physical objects". An later: "experience has shown
\citep{DeForestI} that even automated methods of solar feature tracking, produced by different
authors with the intention of reproducing others' results, have myriad built-in assumptions and
subjectivity of their own unless great care is taken in specifying the algorithm exactly".\\

Motivated by these concerns, in this work we will try to obtain observationally useful
and physically meaningful information about the nature of the magnetic flux concentrations
in the QS, without subjective assumptions about the interaction and
identification of the such features. In particular, we will show that the time-sequence
of the magnetic flux density across surfaces with normal vectors perpendicular to the line-of-sight (referred to as $B_\parallel$),
and normal vector parallel to the line of sight (referred to as $B_\perp$), as well as the magnetic pressure ($B^2=B^2_\perp+B^2_\parallel$), at a
given position on the quiet solar surface verify the properties of a Markov chain. To demonstrate this will employ observations
of the solar magnetic field on the quiet photosphere taken by the {\sc Sunrise}/IMaX
instrument (Section~\ref{section:observations}) and study specific relations for the joint probability and conditional
probability density functions (Section~\ref{section:analysis}) for the three aforementioned time-varying quantities:
$B_\parallel$, $B_\perp$ and $B^2$. The implications of our findings will be discussed in Section~\ref{section:discussion}.\\
\section{Observational data and inference of physical parameters}
\label{section:observations}

The QS data employed in this work has been recorded with the 1-m stratospheric balloon-borne solar observatory
{\sc Sunrise} \citep{Barthol_Imax,Solanki_Imax} with the on-board instrument Imaging Magnetograph eXperiment
\citep[IMaX, ][]{Valentin_Imax}. The data were observed near the solar disk center on June 9, 2009.\\

An average flight altitude of 35 km reduces more than 95\% of the disturbances introduced by Earth's atmosphere, and
image motions due to wind were stabilized by the Correlation-Tracker and Wavefront Sensor
\citep{2011SoPh..268..103B}. IMaX spectropolarimetric data yielded a spatial resolution of 0.25\arcsec and a
field-of-view of 50\arcsec$\times$50\arcsec. Further image reconstruction based on phase diversity calibration of
the point spread function of the optical system improved the resolution to 0.15\arcsec-0.18\arcsec.\\

The IMaX magnetograph uses a LiNbO$_3$ etalon operating in double
pass, liquid crystal variable retarders as the polarization modulator,
and a beam splitter as the polarization analyzer. We use data recorded in the so-called
V5-6 observing mode \citep[see][]{Valentin_Imax}: images of the Stokes vector parameters ${\bf S}=(I,Q,U,V)$
were taken at five wavelengths  ($\pm 80, \pm 40~$m{\AA} from line center, plus continuum at +227\,m{\AA})
along the profile of the spectral line Fe~{\sc i} located at 5250.2 {\AA}. With an effective Land\'e factor of
$g_{\rm eff}=3$ this spectral line is highly sensitive to the magnetic field.\\

\begin{figure}[ht]
\begin{center}
\includegraphics[width=8cm]{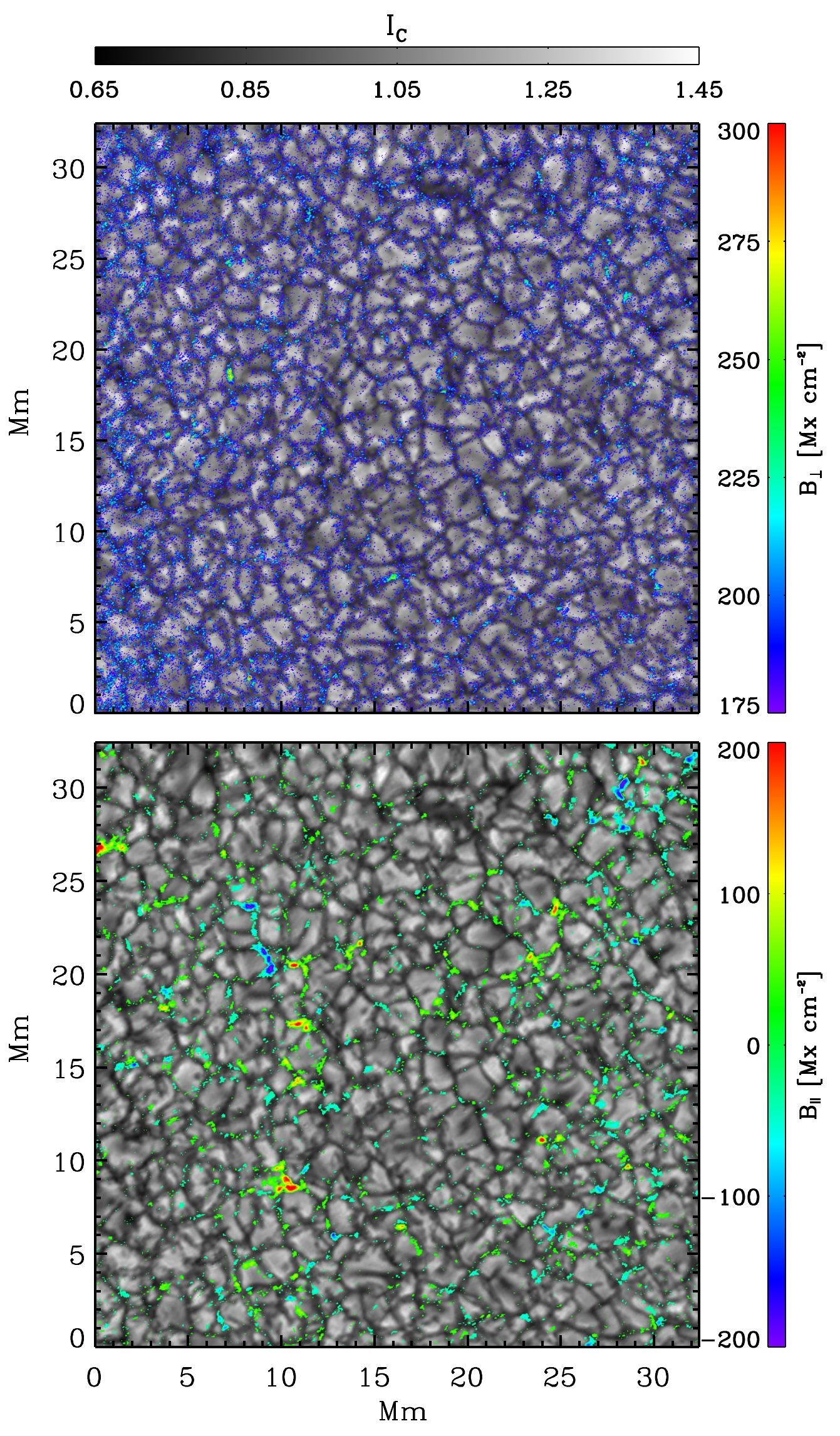}
\end{center}
\caption{Snapshot of the solar surface (i.e. granulation) as seen by IMaX. Gray-scale corresponds
to the normalized continuum intensity. Colors correspond to $B_\parallel$ and $B_\perp$ on the bottom and top panels respectively. Only those pixels where the magnetic flux is at least three times the standard deviation are plotted.}
\label{figure:imax}
\end{figure}

The reduction procedure renders time series of ${\bf S}(\lambda)$ with a cadence of $\T=33$\,sec;
a spatial sampling of 0\farcs055 per pixel, and an effective field-of-view of
45\arcsec$\times$45\arcsec. The total number of available images is $T=113$, yielding a total
observing time of $62$~minutes.\\

From here we infer the longitudinal $B_\parallel$ and transverse
$B_\perp$ magnetic field flux density at each pixel on the detector using an inversion
method based on the radiative transfer equation for the Stokes parameters by means the VFISV code
\cite{borrero2011vfisv}, which assumes that the physical parameters of the atmosphere model
(except for the source function) are constant along the vertical direction in the solar atmosphere within the range of optical-depths where this spectral line is formed (i.e. Milne-Eddington approximation). Following \cite{graham2002}
we refer to $B_\parallel$ and $B_\perp$ as the magnetic flux density through surfaces whose normal vectors
are oriented parallel and perpendicularly, respectively, to the line-of-sight.\\

The signal-to-noise ratio of the observations is affected by: the Poisson
photon noise of the instrument, accuracy of
the polarimetric calibration and quantum efficiency of the detectors.
Following \cite{borrero2011noise} we have estimated a standard deviation
of components $\sigma_{\parallel}\approx 8$ Mx~cm$^{-2}$ and $\sigma_{\perp} \approx 55$ Mx~cm$^{-2}$ as a measure of
our accuracy in the determination of the magnetic field density components.\\

Figure~\ref{figure:imax} shows a snapshot of the solar surface (i.e. quiet Sun granulation)
as seen by IMaX. We also overplot the retrieved values of $B_\parallel$ (bottom panel)
and $B_\perp$ (top panel, but only in those pixels where the inferred values are about three times
above the standard deviation: $|B_\parallel| \gtrsim 25$ Mx~cm$^{-2}$ and $B_\perp \gtrsim 175$ Mx~cm$^{-2}$.

\section{Data analysis}
\label{section:analysis}

In this section, we present a brief theoretical overview on Markov random variables (Sect.~\ref{subsection:theory})
and demonstrate how Markov property has been analyzed and confirmed in our observational data (Sect.~\ref{subsection:test}).

\subsection{Markov property: theory}
\label{subsection:theory}

\begin{figure}[ht]
\includegraphics[]{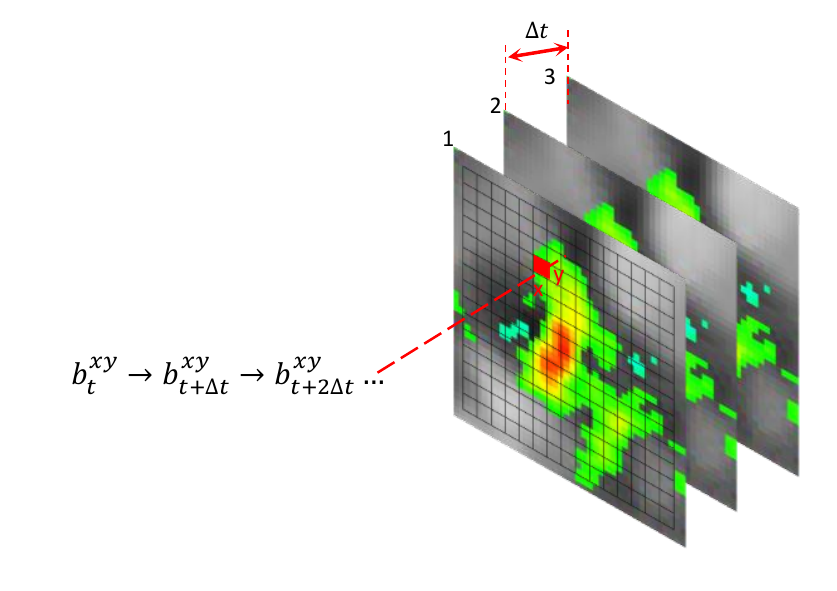}
\caption{Geometry (schematic, not to scale) of the analyzed fluctuations. Image pixels (red square) define a spatial uniform grid at whose nodes we register occurrences (above the noise level) of the observable $\ov_{t}$ in time (see \eref{equation:tran}). The sequences (chains) at each spatial pixel have finite lengths due to interruption by noise and apparent motion of the magnetic concentrations. For the Markov property test, all chains we part into time-ordered pairs and triplets of the random samples (see \eref{equation:p3}, \eref{equation:CK} and \fref{figure:scheme2}). All pixels are considered to be spatially independent contributors across the entire field-of-view to the single set of the registered chains.}
\label{figure:scheme1}
\end{figure}

\begin{figure}[ht]
\begin{center}
\includegraphics[scale=0.85]{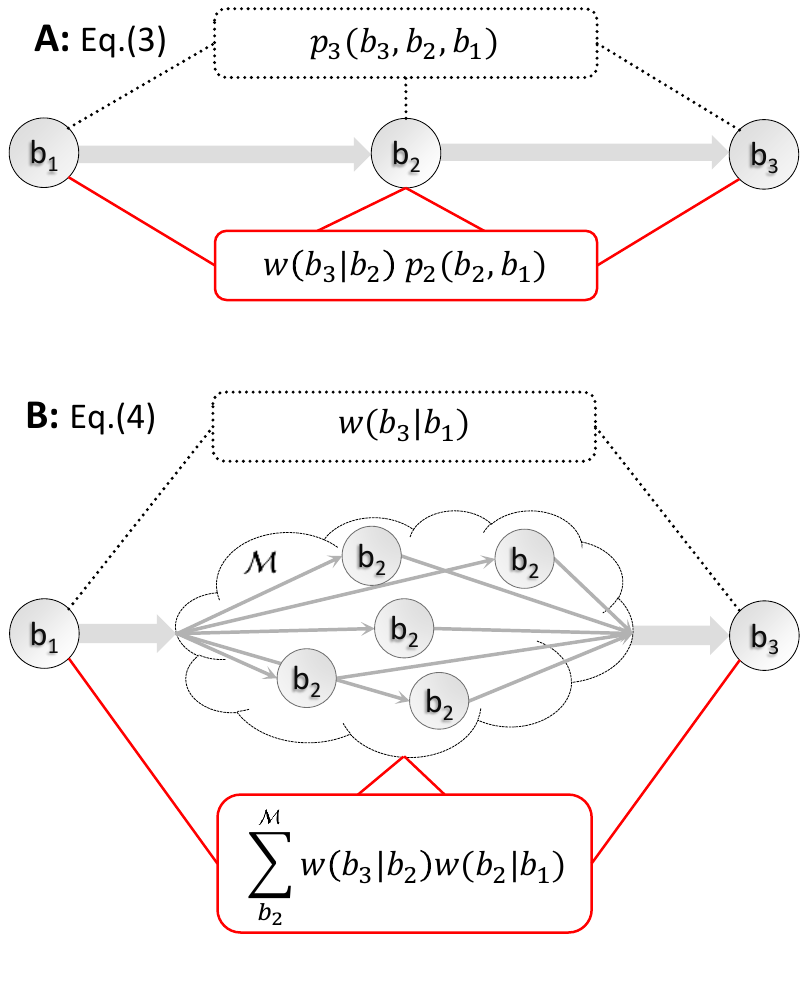}
\caption{Schematic representation of the necessary and sufficient conditions for the Markov property: chart \textbf{A} corresponds to \eref{equation:p3} and chart \textbf{B} to \eref{equation:CK}. Algebraic relations
in \eref{equation:p3}-(\ref{equation:CK}) between statistical quantities are verified by comparison of the independently estimated right-hand side (dotted blocks) and corresponding left-hand side (solid line blocks) of each equation
(see text for rigorous definitions and Fig.\ref{figure:markovfig} for results). Statistic is collected from a set of time-ordered (arrows) pairs and triplets of the random realizations $\{b_1, b_2, b_3\}$ (see \eref{equation:tran} and Section~\ref{subsection:test}), which in turn are acquired as shown in Fig.\ref{figure:scheme1}. Lines connecting formulae blocks and $b_{1,2,3}$-values show dependence of the functions on their arguments. The red blocks designate estimated functions shown with red lines in \fref{figure:markovfig}, and dotted blocks correspond to the circles in \fref{figure:markovfig}.}
\label{figure:scheme2}
\end{center}
\end{figure}

Consider a time-discrete stochastic process $b(t)$, the random variable $b$ is defined over a
finite set of discrete states (state space). The state space has $\mathcal{M}$ distinct elements.\\

Let $\p{n}{b_n ,t_n;\ldots;b_1 ,t_1}\equiv \p{n}{b_n\ldots b_{1}}$ be the $n$-joint
probability density function (pdf) such that $p_{n}(b_n\ldots b_{1} )d^{n}b$ is the probability
that $b$ has values in the interval $[b_1,b_1+db)$ at time $t_1$, \ldots  and in the range
$[b_{n},b_{n}+db)$ at time instance $t_n$. For brevity, the intervals are labeled by the representative
states, that is to say that the process $b(t)$ is in the state $b_m$ at time $t_m$ if the random variable
$b$ has values in $[b_m,b_m+db)$ at time $t_m$. Empirically, $db$ is the fixed binsize that has been
introduced for the estimation of the probabilities, and it is henceforth neglected in the equations for simplicity.
Some trivial properties of the probabilities are $0\le p(b)db\le1$ and $\sum_{b}^{\mathcal{M}} p(b)db=1 $ with $p_{1}(b)\equiv p(b)$.\\

The conditional probability density function $\w{n}{b_{n}|b_{n-1}\ldots b_1}$ is defined such that
$w_n$ is the probability for $b(t)$ to be in state $b_{n}$ at time $t_n$ if the random variable
$b$ already passed through the states $b_{n-1}\ldots b_1$ at later times $[t_{n-1},t_1]$, which we call a
history of the process $\mathcal{H} \equiv b_{n-1}\ldots b_1$. By definition:

\begin{align}
\w{n}{b_n | \mathcal{H} } = \p{n}{b_n \ldots b_1}/ \p{n-1}{\mathcal{H}} \;.
\label{equation:cond-DEF}
\end{align}

A time- and space-discrete stochastic process $b(t)$ is a called Markov chain \citep[\textit{e.g.}][]{Oppen} if
the history of the process $\mathcal{H}$ can be reduced to a single state, which is assigned to be immediately preceding the
current one:

\begin{align}
\w{n}{b_n|\mathcal{H}}=\w{}{b_n | b_{n-1}} \;.
\label{equation:MarkovDef}
\end{align}

It is worth mentioning that $p_{n-1}$ and $p_n$ are functions of $n-1$ and $n$ independent variables being
represented by $\mathcal{M}^{n-1}$ and $\mathcal{M}^{n}$ state configurations, respectively. Due to the moderate
size of the dataset we set $n=3$ in \eref{equation:cond-DEF}
\citep[see also][]{turbulence1997,Friedrich_REV} and obtain the following equation describing the first
condition of the Markov property:\\

\begin{align}
\p{3}{ b_3,b_2,b_1 }= \w{}{b_3|b_{2}}\p{2}{b_2,b_1} \;.
\label{equation:p3}
\end{align}

The second condition we examine is based on the integral form of the Chapman-Kolmogorov equation
\citep[\textit{e.g.}][]{Kampen}, which reflects the time ordering of the chain:\\

\begin{align}
\w{}{b_3 |b_1}=\sum_{b_{2}}^\mathcal{M}\w{}{b_3 | b_2}\w{}{b_{2}|b_{1}} \;,
\label{equation:CK}
\end{align}

\noindent where each $w$ is a $\mathcal{M}^2$ transition matrix.
On their own \eref{equation:p3} and \eref{equation:CK} are necessary conditions for a
stochastic process to have the Markov property, while together they represent also a sufficient
condition \cite[see][and references therein]{PhysRevE.58.919}. Therefore, in the following,
these two conditions are used simultaneously in order to test for the Markov property of the
observed fluctuations in $B_\parallel$, $B_\perp$, and $B^2$ (see \eref{equation:tran}).

\begin{figure*}[ht]
\centering \includegraphics[]{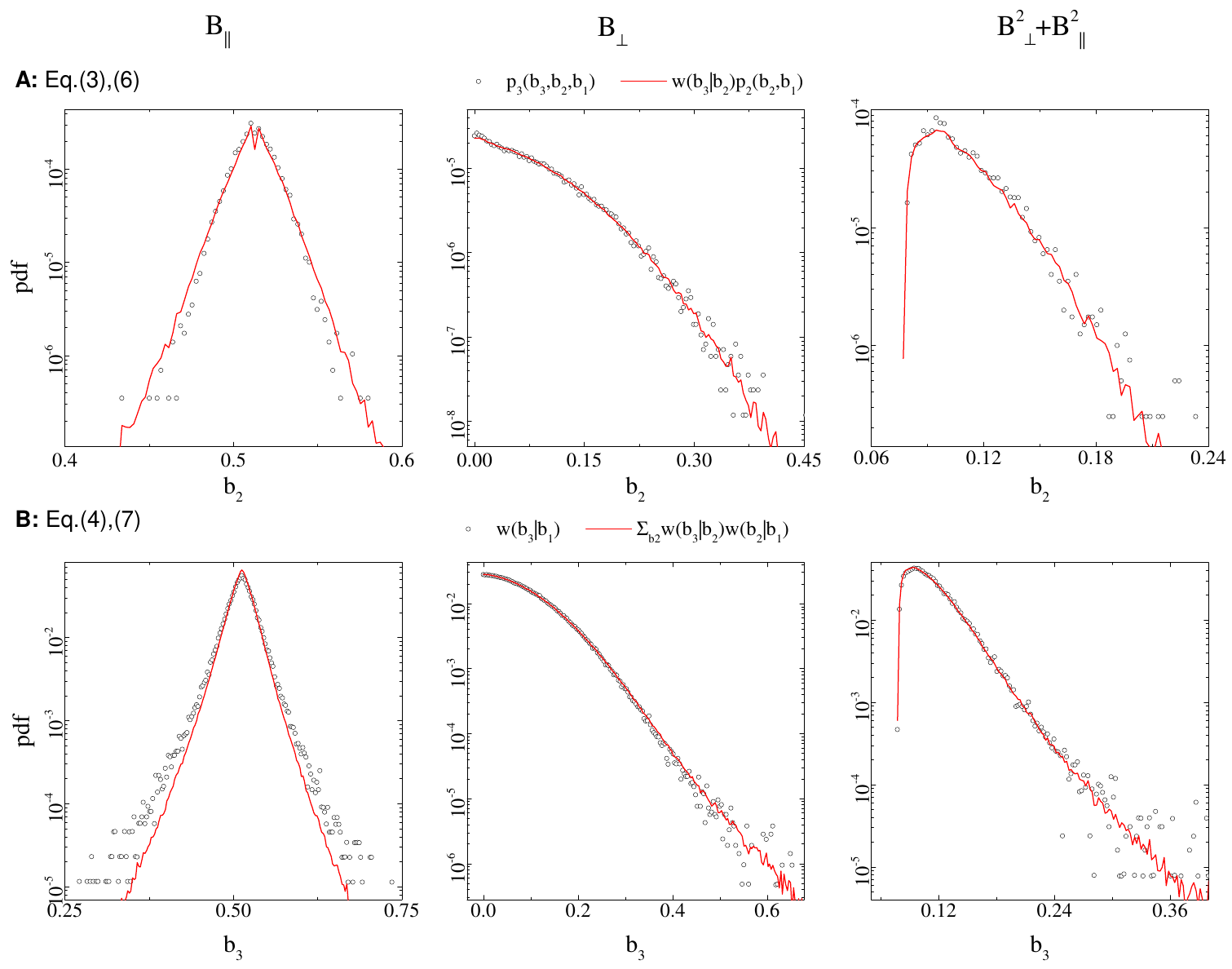}
\caption{The results of the Markov property test. The vertical arrangement of panels corresponds to three
analyzed observables. Top row \textbf{A}: test of the relation given by
\eref{equation:p3avr}. Bottom row \textbf{B}: test of the relation given by \eref{equation:CKavr}.
The abscissa axes are shown in normalized values (dimensionless). The observables $B_{{\sc XXX}}$ are linearly
normalized into continuum interval $b\in [0, 1]$
according to $b=(B_{\sc XXX}-B_{min})/(B_{max}-B_{min})$, where the global extreme
values $B_{min}$ and $B_{max}$ are estimated during the pixel
selection procedure (for double transitions) over all $T$ images. For $B_{||}$, the empty noise cutoff range $[-3\sigma_{||},3\sigma_{||})$
is removed during the normalization procedure.}
\label{figure:markovfig}
\end{figure*}

\subsection{Markov property: test}
\label{subsection:test}

It has been shown by \cite{Ramos_markov} that spatial \textit{increment} $h_r(x,y)=B_{\perp}(x+r,y+r)-B_{\perp}(x,y)$
does not show Markov properties, where $B_{\perp}(x,y)$ is the line-of-sight magnetic flux density
(see \sref{section:observations}) registered at pixel $(x,y)$ and $B_{\perp}(x+r,y+r)$ is the same quantity
but separated by the distance (spatial scale) $r$. In this paper we
perform a similar Markov analysis to the aforementioned work but in the
\textit{time domain} and for the \textit{observables} themselves, not their increments: we examine Markov properties
of transitions/fluctuation of the  observable $\ov_{t}$ in time (from image to image) at a given pixel:\\

\begin{equation}
\ov_{t}\rightarrow\ov_{t+\T}\rightarrow\ov_{t+2\T}\cdots \;,
\label{equation:tran}
\end{equation}

\noindent where $\T$ is the cadence time (\sref{section:observations}), and $\ov$ is one of three
variables ($B_{\perp},B_{||}, B^2$) inferred at image pixel $(x,y)$. {A relation between observable,
image pixel(s) and cadence time is schematically shown in \fref{figure:scheme1}.}\\

{The Markov property is tested by comparing the independently estimated
left- and right-hand sides of \eref{equation:p3} and \eref{equation:CK} (see scheme in \fref{figure:scheme2}).
That is, we count the number of occurrences of the pairs (single transitions) for $p_2(\ov_{t+\T},\ov_{t})$
and $w(\ov_{t+\T}|\ov_{t})$ and triplets (double transitions) for $p_3(\ov_{t+2\T},\ov_{t+\T},\ov_{t})$
according to \eref{equation:tran}. In \fref{figure:scheme2}, the red blocks designate estimated functions shown with red lines
in \fref{figure:markovfig}. The dotted blocks in \fref{figure:scheme2} correspond to the circles in \fref{figure:markovfig}.}  \\

In order to determine the statistics of the transitions described by \eref{equation:tran} we analyze
only those pixels where the signal is above the $3\sigma$-noise cut-off simultaneously at $t$ and $t+\T$
at the same spatial location $(x,y)$. This is done for all conditional probability functions $w$
and the two-joint probability function $\p{2}{b_2,b_1}$ in \eref{equation:p3} and \eref{equation:CK}.
Likewise, for the three-joint probability function $\p{3}{ b_3,b_2,b_1}$ in \eref{equation:p3} the condition is
that the signal must be above the $3\sigma$ cut-off in three images at $t, t+\T$ and $ t+2\T$.
Such pixel-wise analysis of images makes the notion of extended magnetic feature to be irrelevant, as well as
their tracking.\\

The explicit computation of \eref{equation:p3} reveals that the range of values in which $p_3$ is defined,
given by the $\mathcal{M}^3$-dimensional space of independent samples, is quite sparse\footnote{The particular value of $\mathcal{M}$
depends on the binsize: $\mathcal{M}db=1$, whose optimal value is computed as in \cite{knut}. With this, we obtain $\mathcal{M}_{B_{||}}=432,
\mathcal{M}_{B_\perp}=295$, and  $\mathcal{M}_{B^2}=455$.}. {Thus, in order to improve its statistical significance, we select those
triples that have maximal occurrence in the $\mathcal{M}^3$-space and those with occurrence value of at least 90\% of the maximal one.}
 We refer to the set of statistically reliable points as $(b'_{3},b'_{2},b'_{1})$.\\

The test of the Markov property is split into two steps. First, we transform $p_{3}$, $p_{2}$ and $w$ into
$\mathcal{M}$-dimensional vectors by fixing the variables $b_3$ and $b_1$ to each of those points selected
as statistically reliable: $b_{3}=b'_{3}$ and $b_{1}=b'_{1}$:\\

\begin{eqnarray}
\begin{tabular}{cc}
$p_{3}(b_3,b_2,b_1)=p_3(b_2)|_{b'_{1},b'_{3}}$ & \textrm{;} $w(b_{3}|b_2)=w(b_2)|_{b'_{3}}$ \notag \\
$p_{2}(b_2,b_1)=p_2(b_2)|_{b'_{1}}$ & \textrm{;} $w(b_3|b_{1})=w(b_3)|_{b'_{1}}$  \notag\\
$w(b_2|b_{1})=w(b_2)|_{b'_{1}}$ & \\
\end{tabular}
\end{eqnarray}

\noindent such that they transform \eref{equation:p3} into an identity with respect to the free
variable $b_2$:
\begin{flalign}
p_3(b_2)|_{b'_{3},b'_{1}}=w(b_2)|_{b'_{3}}p_2(b_2)|_{b'_{1}}
\label{equation:p3avr}
\end{flalign}
\noindent
and \eref{equation:CK} into $\mathcal{M}$-vector function of the free variable $b_3$.
\begin{equation}
w(b_3)|_{b'_{1}}=\sum_{b_2} w(b_3|b_2)w(b_2)|_{b'_{1}} \;.
\label{equation:CKavr}
\end{equation}

To further increase the statistics, a second step in our test of the Markov property consists in averaging
the left- and right- hand sides of \eref{equation:p3avr} and \eref{equation:CKavr} for all \emph{primed} points
that were previously selected.\\

The results of the described procedure are shown in \fref{figure:markovfig}.
The top row panels in \fref{figure:markovfig} show the estimated relation corresponding to \eref{equation:p3avr} and bottom panels
to \eref{equation:CKavr}. Circles represent the estimated left-hand sides of both equations, while the solid lines
correspond to the right-hand sides. From these figures it can be concluded that, around the global and a few of local maxima of the $\mathcal{M}^3$-space, the Markov property is clearly satisfied.

\section{Conclusions}
\label{section:discussion}

Stochastic Markov processes are intermediate processes that lie between pure randomness of
the independent events and those processes with a strong dependence on the past states (i.e. history)
\citep[\textit{e.g.}][]{Oppen}.\\

Our analysis establishes that the magnetic field temporal fluctuations, as seen by IMaX with a
resolution of 0.15\arcsec-0.18\arcsec and $33$~sec cadence, can be considered as a Markov discrete
stochastic process (Markov chain). The sufficient and necessary conditions for the Markov processes
have been verified for the case of the maxima (global and local) of the available statistics.\\

The revealed Markov property in the temporal dynamics of the turbulent small-scale magnetic field
is the quiet Sun can be used to constraint magneto-hydrodynamics models of the solar atmosphere and a stellar
turbulent dynamo, in general. That is to say, the Markov property should be reproducible in the relevant simulations
of the photospheric magnetic fields.\\

With this work we hope to have brought forward new ideas and techniques for the analysis of solar
spectropolarimetric data. We foresee a number of future applications of the method described in
this paper. For instance, in a future work we plan to investigate the so-called Markov-Einstein time-scale.
This time scale is the minimum time interval over which the stochastic data can be considered as a Markov process.
On shorter time scales, one expects to find correlations and thus memory effects start to play a significant role in
transition probabilities \citep[][and references therein]{Friedrich_REV}. The cadence $\T$ in our
data seems to be greater than (or just equal to) the Markov-Einstein time-scale for the spatial
resolution of our observations. To have an exact relation between temporal/spatial resolution and Markov property,
one needs to perform a systematic analysis of similar observations with different resolutions and cadences.
This will be the subject of a future investigation.\\

This work was supported by the European Research Council Advanced Grant HotMol (ERC-2011-AdG 291659).
We are grateful to Prof. Udo Seifert for useful comments and suggestions. {We thank anonymous referee for valuable comments that helped to clarify and improve the paper substantially.} This research has made use of
NASA's Astrophysics Data System.

{\it Facilities:} \facility{SUNRISE/IMaX}

\end{document}